**Title**: Enhancing the bond strength in meta-crystal lattice of architected materials

**Authors**: M. G. Rashed[a,*], Dhriti Bhattacharyya[b,*], R. A. W. Mines[c], M. Saadatfar[d], Alan Xu[b], Mahmud Ashraf[e], M. Smith[f], and Paul J. Hazell[a]

[a] School of Engineering and Information Technology, The University of New South Wales, Canberra, ACT 2610, Australia

[b] Australian Nuclear Science and Technology Organisation, Lucas Heights, NSW 2234, Australia

[c] School of Engineering, The University of Liverpool, Liverpool, L69 3GH, UK

[d] Department of Applied Mathematics, Research School of Physics and Engineering, The Australian National University, Canberra, ACT 2601, Australia

[e] School of Engineering, Deakin University, Geelong, VIC 3216, Australia

[f] Advanced Manufacturing Research Centre, University of Sheffield, Sheffield, S60 5TZ, UK

[*] Corresponding Author(s): rashed@unsw.edu.au and dhb@ansto.gov.au

**Abstract**: Architected materials produced by powder bed fusion metal additive manufacturing technique offer realization of complex structural hierarchies that mimic the principles of crystal plasticity while still being ultralight-weight, though suffering from deep-rooted multiscale defects including microstructural heterogeneity caused by the complex thermo-mechanical transients in the melt pool. Here we manufacture meta-crystal 316L stainless steel microlattice structures by selective laser melting process for utilizing the strain localization mechanism in bulk structures akin to dislocation slip mediated plasticity. The build angle was observed to be the primary influencer of defects generated and the presence of inherent voids was the major drawback that would undermine their structural performance as mechanical metamaterials. However, other defects in the form of spatially correlated dislocation networks and micro-segregated cellular substructures overcome the strength-ductility trade-off and render the bulk structures comparable to other engineering materials including conventional steels. By exploiting this intrinsic strengthening mechanism, the bond strength of meta-crystals (i.e. strut strength) can be enhanced (or controlled) on top of employing hardening principles of metallurgy to design materials with desired properties.

**Keywords**: Architected materials, Selective laser melting, Mechanical metamaterials, Meta-crystal, Atomic bond strength, Hardening principles of metallurgy.

# 1 Introduction

Metallic microlattice structures are a class of open-cell (cellular) architected materials consisting of a periodic arrangement of unit cells in an interconnecting network of struts and can be realized using additive manufacturing processes, which offer the added advantage of the ease of producing complex geometries without resorting to traditional cutting/tooling approach while still being able to maintain high strength-to-weight ratio. Such hierarchical mechanical metamaterials [1] provide new design and performance attributes for lightweight structures [2,3], though not necessarily free of manufacturing issues [4]. While the additively built microlattice structures reported in literature had low relative density and were found to be similar to those typically reported for other cellular materials [2], the parts were reported to be suffering from internal porosity that would instigate early failure/collapse [4], and deviations in the nominal dimensions were often observed from the target geometry [5]. There appears to be a complex correlation between materials processing parameters such as laser power, laser scanning speed and layer thickness (i.e. build parameters), and the quality and performance of the realized microlattice structures, yet void formation seems to be unavoidable [6]. Despite the inherent porosity and quality issues, the behaviour of Laser Powder-Bed Fusion (L-PBF) manufactured 316L Stainless Steel (SS) microlattice structures has been of particular interest due to their comparable performance (even better in some cases) to other cellular structures composed of lighter alloys, e.g. Al [2], that was puzzling given the apparent high mechanical properties of the parent material in contrast to typical 316L SS and the unit cell topologies utilized. This warrants a detailed investigation on the features and the process-induced imperfections/defects present microstructurally and their effect on the mechanical performance.

Incidentally, the periodic pattern of architected lattice materials has been observed to be analogous to metallurgical crystal microstructure, where the nodes and struts in the unit cells act as atoms and atomic bonds in the crystal lattice, respectively, and the atomic-scale hardening principles of metallurgy (e.g. grain-boundary, precipitation, multi-phase hardening etc.) have been well demonstrated in multiscale hierarchical crystal-like meso-structures [7]. Besides having control over the typical plasticity phenomena in the meta-crystal lattice of architected materials, higher strengths of meta-atomic bonds (force resisted by struts) can be achieved by either increasing the strut diameter or by harnessing the strengthening effect of the parent material in the additive manufacturing process, while the former would cause a density hike, the latter would enhance specific strengths (i.e. better light-weighting) and is the focus of the current study.

This paper addresses additive manufacturing of 316L SS lattice structures by the Selective Laser Melting (SLM) process using a special laser exposure technique (i.e. scanning strategy) that enabled realization of thin strut sections at the micro-scale. Furthermore, the paper reports in-depth analysis of the quality of the as-built structure, the microstructure of the parent material, and respective mechanical properties for a specific set of lattice configurations (a primitive unit cell against a meta-crystal unit cell, that act as control group and experimental group, respectively, to single out the independent variable: meta-crystal behaviour). Individual struts (build angle of 35.26° from BCC and 90° from BCC-Z unit cell, Supplementary Fig. 1), that represent atomic bond in meta-crystals and the smallest/fundamental building block of each unit cell (i.e. macro-crystal), extracted from manufactured microlattice structures were investigated using various microscopy techniques, mechanically characterized, and the question of the surprisingly high performance is elucidated by correlating the melt pool thermodynamics with the microstructural observations. This paper has thus endeavoured to develop the understanding of the complete fabrication process, hence controlling and ultimately optimizing the process, to realize multiscale hierarchical (to mimic multi-grain, and/or precipitates) lightweight structures in typical crystal lattice configurations or optimized unit cell topologies (lone, or in a combination to mimic multi-phase) tailored for specific use case scenarios.

## 2 Materials and Methods

### 2.1 Design

The microlattice structures of cubic block (20×20×20 mm$^3$) manufactured by SLM process in this research comprise two different periodic unit cells (of different side lengths: 2.50, 2.00, 1.50, and 1.25 mm) arranged in a repetitive order (Supplementary Table 1): (a) Body Centred Cubic (BCC) and (b) the same unit cell with an additional vertical member at the centre (BCC-Z) (Supplementary Fig. 1a). The discrepancy between the depiction of unit cells and the actual definition (Supplementary Fig. 1b) was implemented to ensure that there is at least one complete (full diameter) vertical strut in every unit cell (macro-crystal) of the BCC-Z structure (Supplementary Fig. 1c), and the BCC structure followed suit to maintain conformity. Additionally, the unit cells did not have horizontal struts, unlike in atomistic crystal lattices since (currently) this build orientation results in manufacturing failure of struts in single spot exposure technique (build angle must be ≥ 25° to the horizontal [2], though horizontal strut would be achievable in the foreseeable future as the technology matures), and vertical struts connecting the unit cell corners were also omitted to maintain uniformity in principal directions (except for BCC-Z to mimic dislocation slip occurrence, i.e. meta-crystal). Although the BCC-Z unit cell has equal edge length in all three orthogonal directions (X, Y and Z), it is not strictly a "cubic" meta-crystal, because the 3-fold symmetry along the cubic body diagonal which is characteristic of cubic lattices is not present as the horizontal struts are absent (that would have acted as atomic bonds, thus weaker in X and Y directions). Therefore, there is a 4/mmm tetragonal symmetry (I4/mmm space group), and it mimics a crystal lattice close to Body-Centred Tetragonal (*bct*) such as Martensite. Since the blocks contain uniformly orientated unit cells, they would work as single meta-grains in the context of meta-crystal structures [7].

## 2.2 Materials and Manufacturing

LPW Technology® (UK) supplied the AISI 316L SS powder manufactured by nitrogen gas atomization, which consisted of well-defined spherical particles with varying size (improving the packing efficiency) and followed a normal distribution with average particle size of 38 μm (Supplementary Fig. 2a and b). The composition of 316L SS powder was obtained by different chemical analysis techniques: Inert Gas Fusion (IGF) for O, N and H; combustion infrared detection for C and S, and ICP-OES for rest of the elements; this was compared with the manufacturer's specification and the composition was found to be within the specification limits (Supplementary Table 2). All the 316L SS microlattice structures were manufactured in MCP Realizer II SLM™ machine by continuously depositing powder layers of 50 μm thickness and the single spot scanning strategy was used as the target strut diameter was ~200 μm, then a laser spot size of 90 μm (with 140 W power and a "single" exposure of 500 μs) resulted in a melt pool diameter equal to that of the desired strut (Supplementary Table 3).

## 2.3 Materials Characterization

### Nano-CT

Individual struts were collected from the lattice structures by wire-cut Electrical Discharge Machining (EDM) to avoid inducing stresses during the extraction procedure to conduct tomography, electron microscopy and nanoindentation experiments. The nano-scale computed tomography (Nano-CT) data acquisition was done in an in-house developed helical CT scanner where individual 316L SS struts placed vertically in a hollow strut holder were exposed to 100 keV X-ray energy for 6 seconds at a sample-to-detector distance of 10.896 mm. The sample was rotated in a helical fashion (combination of vertical and rotational travel) during the data acquisition phase, and typically took ~24 hours to finish. The temperature and humidity of the data acquisition room were carefully controlled, to avoid nano-scale expansion or contraction, and compensated in both day and night-time to maintain constant climate parameters during the entire experiment. Finally, the iterative reconstruction procedure yielded images at a resolution of 411.602 and 428.113 nm for the 35.26° and 90° strut, respectively.

*Electron Microscopy*

A Zeiss® UltraPlus™ SEM Equipped with Oxford Instruments® Electron Back-Scatter Detector (EBSD) and Energy Dispersive Spectroscopy (EDS) detector was used to collect orientation and elemental data from the struts with the detector tilted at 70° for EBSD data analysis [8]. All SEM data were collected at 20 keV. Transmission Electron Microscopy (TEM) was conducted in a Jeol® 2200 FS™ S/TEM instrument operated at 200 keV and equipped with Oxford Instruments® EDS detector. TEM thin foils were prepared from the respective struts by using a Zeiss® Auriga 60™ Cross-beam™ Focused Ion Beam (FIB) instrument with a Ga ion source operating at 30 keV.

*Dislocation Density Quantification*

TEM micrographs were taken from different areas of each foil and the linear-intercept method was used for dislocation counting, where each micrograph was sub-divided into square grids (16 or 9 squares), improving the statistical homogeneity of the dislocation counting result. Each square was analysed separately within the image by drawing a number of random lines (6 to 7 in this study) across the square, and the intercepts of the dislocations with the lines were counted. The dislocation density ($\rho$) in each square was calculated [9] by,

$$\rho = \frac{2N}{Lt} \tag{1}$$

Where, $N$, $L$ and $t$ represents the number of intercepts, summation of line length and thickness of the foil, respectively. The thickness of the foils was estimated to be 150~200 Å. Finally, the mean dislocation density was calculated by averaging the density in all the squares in micrographs of respective struts, followed by invisibility corrections for reflection vector **g**.

## 2.4 Mechanical Characterization

*Strut Nanoindentation Tests*

An Hysitron® TI 900™ TriboIndenter equipped with a diamond Berkovich tip was used to measure hardness of the as-built struts. The (resin mounted) specimens were indented perpendicularly to their respective surfaces, using a maximum indentation load of 10 mN, a loading time of 10 sec and unloading time of 10 sec. A 3×4 grid of (total 12) indents was made for each sample with 20 μm and 30 μm spacing in both X- and Y-directions, to ensure maximum number of indents in different grains, for inclined and vertical strut, respectively.

*Bulk Compression Tests*

Compression tests on the bulk lattice cubes were carried out in an Instron® 4024 Universal Testing Machine™ (UTM) as per ASTM E9-89a (2000) [10]. The top and bottom platens were greased before each test to eliminate friction between the platen-lattice interface. The cubes were loaded in the build direction (along the vertical pillars of the BCC-Z lattices) at a quasi-static strain rate ($0.00042$ $s^{-1}$) where the displacement was determined with an extensometer attached to the upper and lower compression platens.

## 2.5 Finite Element Analysis

Compression of both unit cells (of 2.50 mm sides) up to 80% of strain at the same bulk lattice structure quasi-static strain-rate was simulated in Finite Element program ABAQUS. Boundary conditions were applied to the top, bottom and sides of the unit cells (by using coupling constraint equations to keep the surfaces straight for mimicking a unit cell constrained by other surrounding cells). Additionally, the bottom was fully fixed and the top was restrained in all directions but vertical to allow downward compression displacement. A general purpose reduced-integration hexahedral element C3D8R was used in the mesh (BCC: 76,236 and BCC-Z: 94,380 elements in total) with isotropic elastic–plastic material behaviour taken from nanoindentation experimentation and Kashyap et al. [11]. Finally, a node-to-surface contact algorithm was used to account for interaction among the struts during compression, where a 'hard' pressure-overclosure (normal) and penalty friction coefficient of 0.55 (tangential) behaviour was defined.

## 3 Melt Pool Thermodynamics and CALPHAD

The SLM process involves extremely high cooling rate ($10^3$-$10^8$ K/s) [12], where the combination of a small laser spot size and high laser power focused on a patch of powder surface produces a very high energy density, and the resultant heat would be transferred either to neighbouring solidified/semi-solidified regions by conduction and/or to the atmosphere by radiation or convection from the top of melt pool (Supplementary Fig. 3a) [13]. Subsequently, the combination of the surrounding build compartment temperature, already solidified metal and fresh metal powder may act as a heatsink to cool the melt pool much faster than it would be in air or even in water (i.e. apparent quenching), resulting in immediate crystallization of the melt pool (Supplementary Fig. 3b). The laser beam would travel back on top of a previously melted spot and melt a new layer of freshly deposited powder, thus inducing re-heating and partial melting in zones underneath, creating a distinct Fusion Zone (FZ) and a Heat Affected Zone (HAZ), akin to welding solidification [14]. The laser beam then moves to the next spot on the horizontal 2D slice before coming back again on a subsequent upper layer, leaving a time delay in revisiting vertically aligned spots that enables the melt pool in the bottom layer to be solidified more than that of the immediate upper layer. Hence, the layer-by-layer build paradigm and resultant through-thickness variational solidification process (Supplementary Fig. 3c) is part of the focus of the current study.

CALPHAD based thermodynamic property calculations were performed to evaluate the melt pool thermodynamic behaviour considering the chemical composition of 316L SS powder, using the Fe alloys database from Thermotech Ltd. [15], which yielded a density of 7867.61 kg/m$^3$ at room temperature (Supplementary Fig. 4a). The rapid cooling rate comes into effect to achieve fully austenitic microstructure immediately after solidification finishes, possibly between 1200 to 1000°C (Supplementary Fig. 4b). Presence of any other phases beside austenite is denied, as none of them could start transformation (0.01%) (Supplementary Fig. 4c) due to the high cooling rate (> $10^5$ °C/s [16]). The elemental composition at equilibrium (Supplementary Fig. 4d) in the 1200 to 1000°C temperature range matches that of chemical analysis, resulting in the rapidly cooled SLM 316L SS to have ~100% austenitic (*fcc*) microstructure [17], with trace amount of MnS at the same temperature range (Supplementary Fig. 4e).

## 4 Results

### 4.1 Microstructural Defects

The 3D nano-CT volume data analysis reveals the shapes and sizes of the internal porosity of both inclined and vertical struts in Fig. 1a. The size (equivalent diameter) and shape (aspect ratio) distribution data of the voids present inside the strut is shown in log-normal plots to clearly distinguish lesser occurrence frequencies. The defect distribution was found to be build angle dependent - the vertical build angle produced comparatively larger sized voids (high volume fraction) and aspect ratio than the inclined build orientation. The void volume fraction was quantified by dividing the sum of the volume of all voids with the volume of the solid strut and estimated to be 0.724% and 1.220% for the inclined and the vertical strut, respectively. Both build orientations have high occurrence in 0~5 μm range, which is understandable as the said range includes sub-micron (nano-meter) scale voids as well that are prevalent in both samples; while the aspect ratio stay mostly within ~1-2 for both. The widely reported stair-case surface morphology for inclined build angle is apparent [18]. As a result, the surface roughness would be higher for samples with inclined build angles, though relatively high surface roughness is a by-product of L-PBF process [19], irrespective of the build angle.

The trend of grain growth is observed to be directed somewhat parallel to the building direction from the EBSD analysis result shown in Fig. 1b, with the 90° strut exhibiting columnar grains of large size, whereas the 35.26° strut has mixed regions of columnar (larger) and equiaxed (finer) grains, i.e. more equiaxed than the vertical strut. Similar preference has been observed in L-PBF *fcc* metals for grain growth [20]. A propensity of <001> crystallographic texture along the build direction was observed for the vertical strut as it has been typically reported [21] and is the easy/preferred growth direction for *fcc* metals [22], while the inclined strut exhibited a different trend. The grain size was 17.17 μm and 39.42 μm for the 35.26° and 90° strut sections, respectively.

The BSE images in Fig. 2b show the presence of cellular structure, and, in some regions, substructure in the as-built samples, the "cells" being grains/sub-grains formed during solidification. A substructured grain microstructure was more prominent in the vertical strut than in the inclined strut (Fig. 2a), though regions with no clear substructure were also observed throughout both samples. Inclusions accumulated at the intersections of cell walls are observed in the substructured region of the vertical strut (i.e. solid particles present in the melt that get trapped as the material solidifies [23]) along with the absence of tree-like structure (dendrite arms), which is evidence of cellular solidification. A random distribution of inclusions (at grain boundaries and interior) was seen in the non-substructured regions.

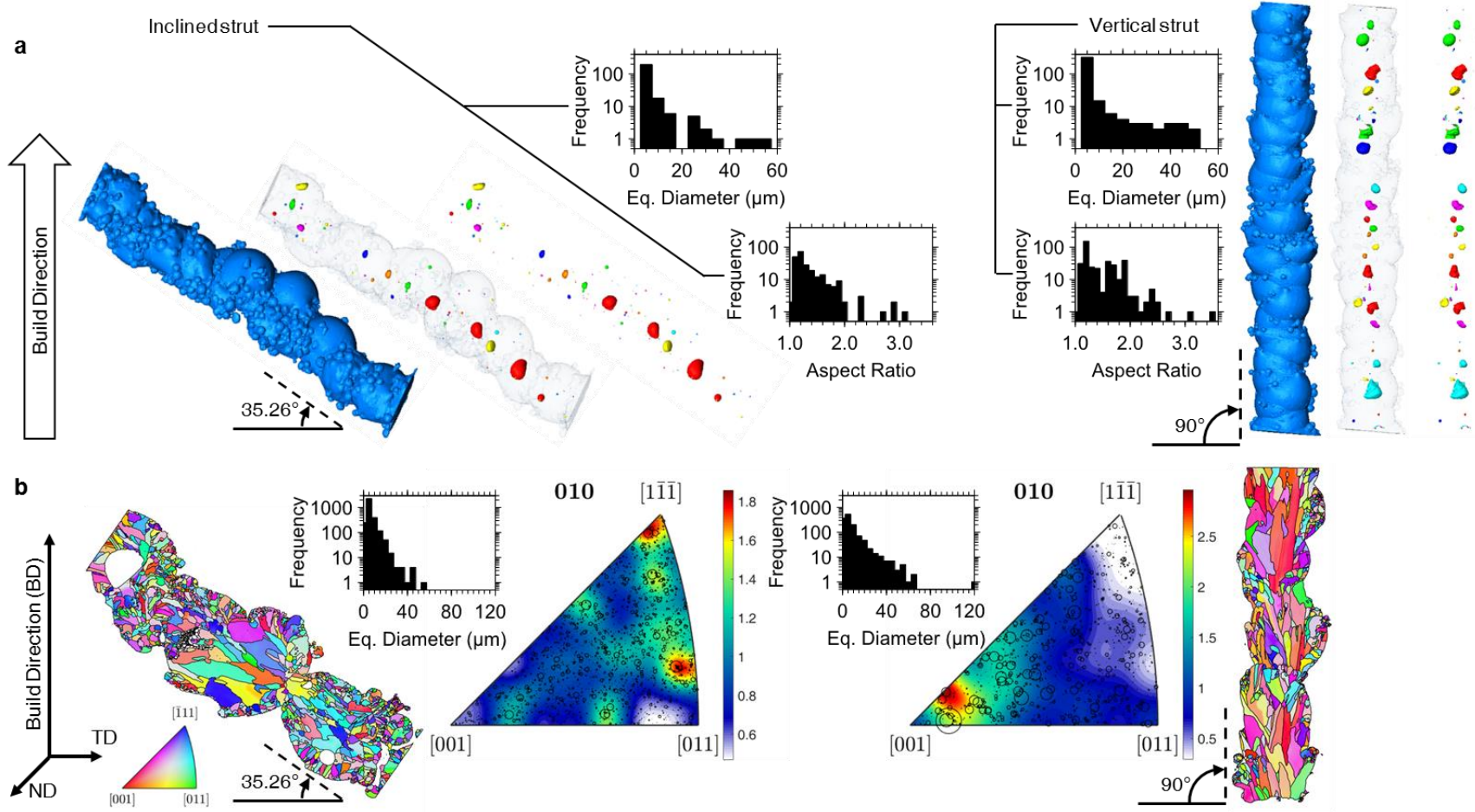

Figure 1: (a) 3D nano-CT scan reconstructed image, spatial position and distribution (size and shape) of voids in both struts, (b) EBSD analysis result of strut longitudinal-sections: IPF maps (boundaries ≥ 15°) represents the orientation distribution with respect to the build direction (the sector represents the colouring scheme) and IPF plots in the same direction (circles represent grain size).

Detailed TEM analysis revealed the presence of three types of dislocation structures – (i) cellular, (ii) ladder-like, (iii) individual segments. Fig. 2c shows cellular and ladder-like dislocation structures, while Figs. 2d and e show the heterogeneity of cellular wall characteristics (e.g. cell formation stages, size, width, dislocation density etc.), bowed-out individual segments, cell wall formed by entangled dislocations, and dislocation ends pinned by precipitates (inclusions), respectively, in a hierarchal length-scale. The solidification substructures in this study align with the dislocation cell structure. The former should pre-exist before the dislocation network forms from a traditional point of view, since they are defined chemically as a diffusional effect (artefact) of solidification, while the latter is the result of complex thermo-mechanical phenomena in the powder bed during laser beam fabrication. The 90° samples seem to have a higher dislocation density than the 35.26° sample. From the manual linear-intercept method used on TEM images, a dislocation density of $3.51 \times 10^{14}\ m^{-2}$ and $8.84 \times 10^{14}\ m^{-2}$ was obtained for 35.26° and 90° strut, respectively.

Both samples appeared to have randomly distributed spherical nano-sized (< 50 nm) inclusions (aforementioned), a representative dispersion of which is shown in Fig. 2f. EDS data acquisition for line profile in STEM mode revealed they are hard ceramic (non-metallic) inclusions [23] relatively rich in Si, O and Mn compared to the matrix, as shown in Fig. 2g. They are solid even when the metal is fully molten [23] and get trapped in between the solidifying cells as the liquid cools, causing them to be located at cell walls (the last part to solidify). Typical MnS precipitates, predicted by CALPHAD calculation and often reported in conventionally manufactured 316L SS [24], were replaced by Si- and Mn-oxides due to in-situ oxidation during laser melting at high temperatures in presence of residual oxygen in the build chamber [24,25], and subsequently retained in the microstructure through rapid (non-equilibrium) solidification.

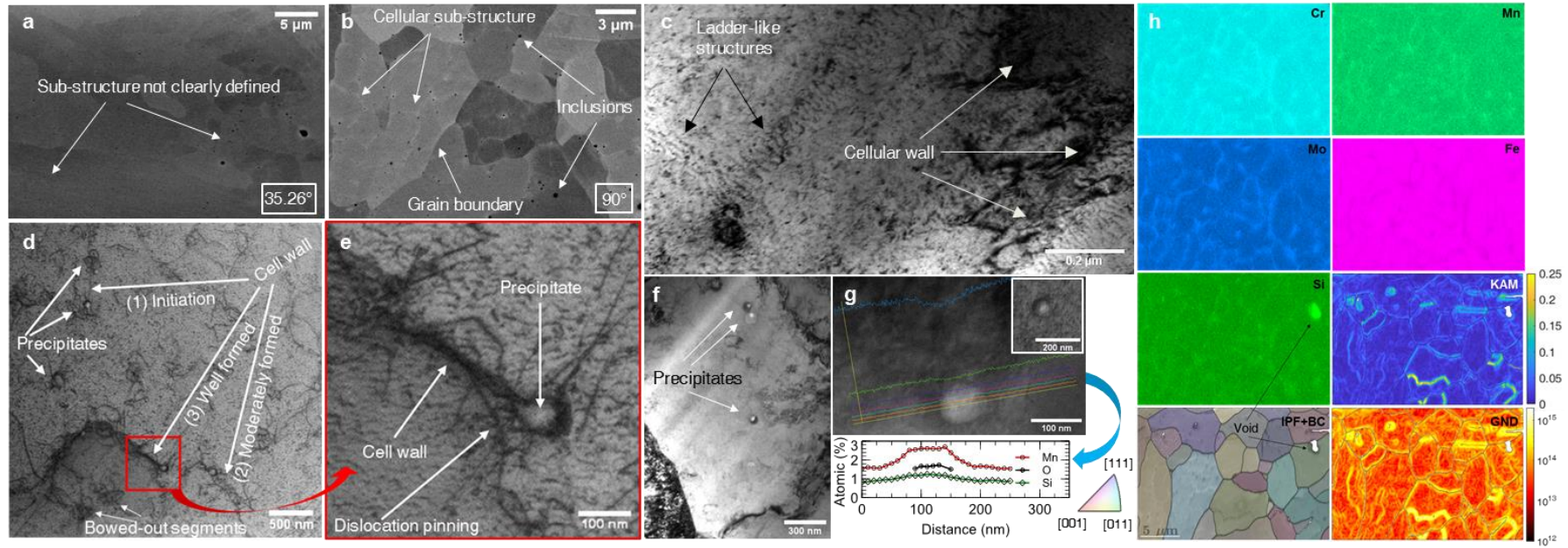

Figure 2: As-built sample images: (a-b) presence/absence of cellular substructure in BSE, (c) dislocation structures in TEM-BF, dislocation cell wall in STEM-BF - (d) stages of cell wall formation, and (e) the pinning mechanism, (f) presence of precipitates in TEM-BF, (g) STEM-EDS shows precipitates (TEM-BF inset) that are rich in Si, O and Mn, (h) SEM-EDS (step size 50 nm) elemental maps show segregation to cell walls and depletion of Fe therein accompanied by IPF shaded by Band Contrast (enhanced), KAM and GND map.

As Fig. 2b (BSE image) show elemental contrast between the grain boundaries, grain interiors and cell walls (substructured regions); segregation was evident in the SEM-EDS elemental maps shown in Fig. 2h. Higher solute contents of Cr, Mo and Si, along with traces of Mn, were observed to be deposited on the substructure walls that caused depletion of Fe. On the contrary, segregation of these elements or other(s) were not detected in the non-substructured regions (Supplementary Fig. 5).

**4.2 Mechanical Behaviour**

The bulk compression test result for lattice blocks with four (4) different unit cell sizes (edge lengths) are shown in Fig. 3a, with nanoindentation hardness maps (Fig. 3b) for extracted individual struts (inclined from BCC and vertical from BCC-Z). The vertical pillar in BCC-Z appeared to be strengthening the response compared to BCC, and the general trend of smaller size unit cells (higher relative density, hence greater rigidity because of more junctions/nodes and struts per unit volume) performing better is seen for both types of unit cells. The BCC-Z blocks emulate typical *bct* single crystal deformation (micropillar compression in Fig. 3e) from the nature of stress-strain curve and slip-like shear band formation [26]. This meta-crystal lattice shears along one or more preferred nodal planes (Supplementary Fig. 1b) upon yielding by having expanding shear bands in the deforming block (Fig. 3d), which was absent (i.e. uniform deformation) for BCC blocks (Fig. 3c). In addition, the hardness of as-built struts is observed to be rather high compared to annealed 316L SS hardness of 1.96 GPa [27], which indicates enhanced yield strength. The intrinsic (void-free) yield stress (0.2%) of the inclined and vertical struts were calculated to be ~545 and ~641 MPa, respectively (Supplementary Calculation 1, [28,29]), compared to the typical yield stress of 205 MPa reported for conventionally manufactured 316L SS (annealed) [30]. Similar high yield stress has been reported in existing literature [16,31,32] for bulk samples with varying degrees of porosity.

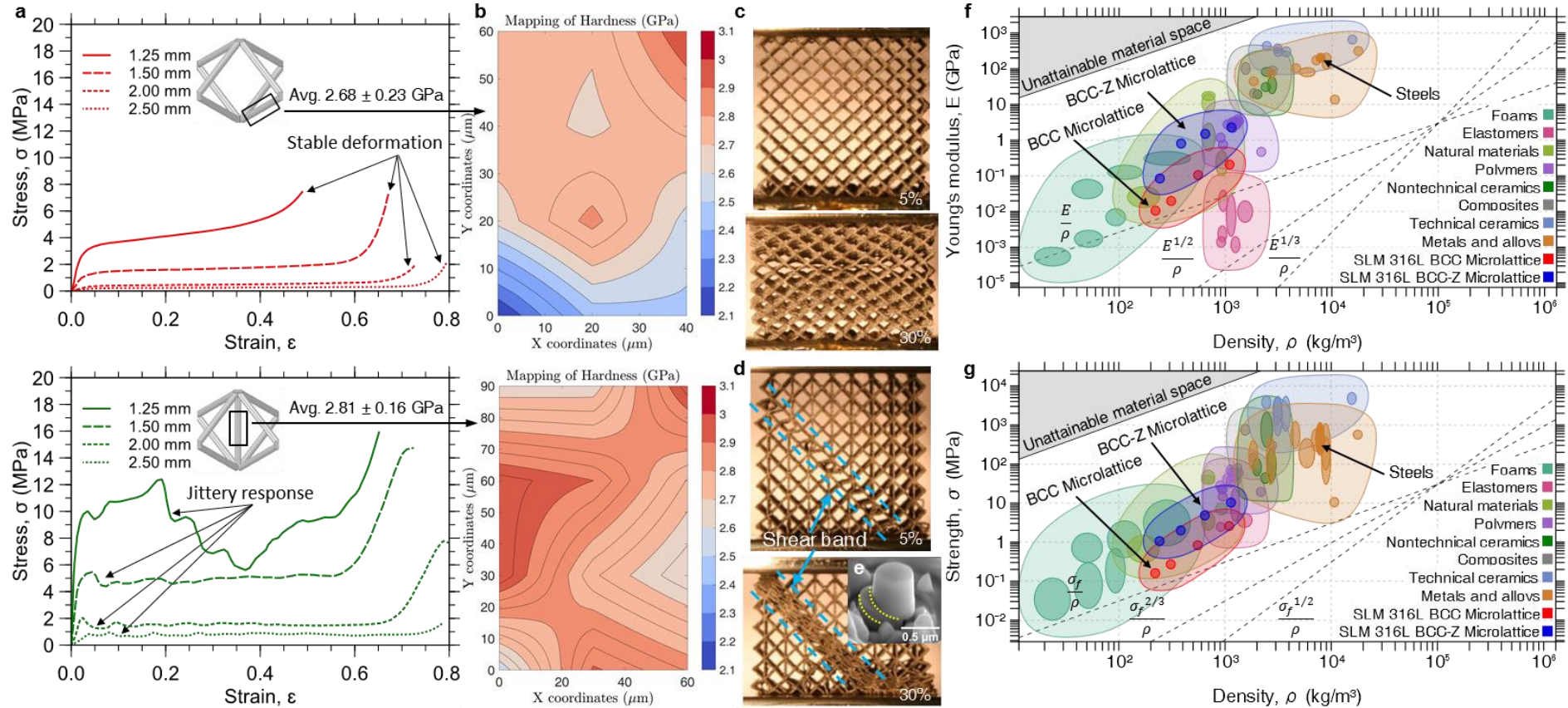

Figure 3: (a) Compressive stress-strain response of both BCC and BCC-Z unit cell microlattice structures ($20^3$ mm$^3$) up to 80% strain with four (4) different unit cell sizes, (b) nanoindentation hardness plot of respective dominant strut (mid-section along the length) exhibits higher avg. hardness (H) than annealed 316L SS, (c and d) deformation mode of BCC and BCC-Z blocks (unit cell with 2.50 mm sides) at 5% and 30% strain, respectively; (e) slip in a martensite (*bct*) single crystal (dotted yellow line is slip trace) [26], Ashby plot on (f) Stiffness vs Density, and (g) Strength vs Density.

From the Ashby plots of Young's modulus and Strength vs Density (material selection charts plotted from bulk compressive properties in Supplementary Table 4, among additional data for common engineering materials [33]) in Fig. 3f and g, the performance of both microlattice structures (relative density of 2.80 to 13.88% for BCC, and 3.07 to 14.45% for BCC-Z) are observed to be comparable to conventionally manufactured steels (100% relative density) enabled by the high yield strength of the base/parent material (i.e. SLM 316L SS), especially for the meta-crystal BCC-Z structures; notably for bending load and surpassed only by the trio of light metals (Ti, Al and Mg alloys).

## 5 Discussion

### 5.1 Contribution of Powder-Bed Mechanism

#### 5.1.1 Dislocation Structures

The process of stress direction reversal (expansion and/or contraction against constraints) in the melt pool during manufacturing acts as pseudo-cyclic loading because of successive re-visitation of the laser beam on top of previously fused spots to melt a fresh powder layer and causing thermal expansion (Fig. 4a). The melt pool cools rapidly upon the exit of the laser beam, leading to thermal contraction in the fusion zone (FZ) where tensile stress would develop. As a result, surrounding regions in the immediate vicinity would develop compressive stress [34]. The comparative heating-cooling temperature rate (Fig. 4b) and the ensuing successive loading-unloading cycles (Fig. 4c) generates the rather peculiar feature of ladder-like dislocation structures (i.e. fatigued samples [35]), where the continuous reversal of stress direction causes – (i) dislocation dipoles to form, and (ii) Persistent Slip Bands (PSBs) to evolve. Such localization of the cyclic plastic strain to the PSBs (formation of ladder-like structures) has been reported extensively for *fcc* metals/alloys [36], including 316L austenitic SS [37] at elevated temperature [38], even after only a few thermal cycles [39].

Similarly, cellular dislocation walls form in PBF due to inhomogeneous heating/cooling that results in nonuniform deformation (thermal strains) by expansion/contraction (i.e. Taylor forest hardening [11]). The residual stresses exceed the yield stress [40] during the build process as a consequence of significantly lower yield stresses at elevated temperatures [41], and produces well-defined cell walls by either the entanglement of generated dislocations with precipitates/inclusions (Fig. 2d) or the entrapment of dislocations in a region with a higher solute concentration (because of the elastic strain fields of segregated atoms). The third observed dislocation structures, consisting of bowed out dislocation segments, is formed by fine inclusions, which would cause dislocations to get pinned (Fig. 2e) and bow them out (Frank-Read mechanism).

**5.1.2 Substructures and Grain Growth**

Two different substructures were discussed earlier, thermal deformation originated dislocation network and solidification induced cellular substructures. Which one forms first has been a topic of debate [42] although both appear to be spatially coincidental, with reports of intermittent presence of substructures in contrast to frequent dislocation structures, observed in this study and by others [42]. The substructure formation would rely mainly on the solidification process that may trigger planar or elemental segregation induced non-planar solidification front (cellular or dendritic) depending on the thermal condition such as the ratio of the thermal gradient (G) to the solidification rate (R) and undercooling ($\Delta$T) [23,43], with the former solidification mode resulting in no chemical heterogeneity and the latter exhibiting substructures. The inclined build angle had each successive layer off-set horizontally that necessitated a course change in the laser travel path for each new build layer, facilitating a faster solidification rate and larger thermal gradient (G) vertically across the melt pool (solidification direction, normal to the solid-liquid interface) as a result of a lesser extent of re-visitation (laser shifted away horizontally in subsequent visitations after each single exposure, and the ensuing fast heat transfer mechanism will cause a smaller HAZ). Contrarily, the vertical build angle had successive melted layers directly on top of each other without any offset in the laser spot, therefore the layers underneath were not able to go through rapid solidification permanently due to the continuous re-visitation of the laser (i.e. thermal cycles) and subsequent re-heating (larger HAZ), which slowed down the effective solidification rate and maintained a smaller thermal gradient. This would cause the inclined build orientation to have higher G/R than the vertical, effectively moving the solidification mode to the top-left on a G-R plot [44], similar to observations made in fusion welding [45]. Consequently, the vertical build orientation would primarily experience cellular solidification and substructure formation (evidenced by BSE image in Fig. 2b), while the deviation of strut build orientation from vertical to inclined would shift the dominant solidification mode from cellular to planar, and mixed mode solidification in between. Similarly, the grain size depends on the combined effect of thermal gradient and solidification rate (G×R, i.e. cooling rate) - the smaller the gradient and slower the rate, the coarser is the grain size [14]. Hence, the grain morphology observed in the struts (Fig. 1b), though both show an affinity to form finer equiaxed grains near the regions where there is a greater free surface area available for rapid heat dissipation (exteriors, near voids, possible porosity underneath the surface or existed in the immediate vicinity of the polished out portion). In addition, the build paradigm of inclined strut (i.e. faster solidification rate) increased required diffusion distance, thus prohibiting development of <001> crystallographic texture along the build direction because of the misalignment [14] between solidification (vertical) and heat flow direction (along strut axis). The overall effect of thermal cycles on microstructure with respect to build angle is depicted in Fig. 4d.

Until now, the formation of cellular shaped substructure in SLM process has been attributed to solute segregation at the cell boundaries [17,23,25] as a result of non-equilibrium cooling conditions that prevent diffusion from occurring fully in the solid, accompanied by solid-state partitioning of some ferrite stabilizers (Mo, Cr and/or Si) [23,46,47]. Based on the liquid-austenite phase transformation partition coefficient of solutes in this study (Supplementary Table 5), similar trends ($k^{\gamma/L} < 1$) have been observed for Cr, Mo and Mn (Fig. 2e), yet that did not explain the segregation of Si ($k_{Si}^{\gamma/L} \sim 1$) in the substructured regions and the "prominence" of dislocation structures ($\rho = 3.51 \times 10^{14}\ m^{-2}$) in the non-substructured regions (i.e. in the absence of solute enrichment) that appear to be beyond the sole contribution of thermal stresses (Supplementary Calculation 2, [25,11,48,49]). It is suggested that akin to the "chicken or the egg" situation where one pre-existing condition would induce the other, cellular substructures would impede dislocation motion by slowing them with segregated solute atoms/particles (i.e. dislocation trapping) [50] and dislocation networks would form substructures by solid-state solute diffusion (aided by the formation of Cottrell atmospheres) in a process similar to Dynamic Strain Aging (DSA) [42], or both mechanisms would be in effect simultaneously to a varying degree, with the pre-existing condition acting as the seed for the end geometrical form (i.e. cellular) in both cases. Austenitic stainless steels show diffusion driven solute segregation through DSA in a temperature range from 473 K to 1073 K (200 °C to 800 °C) [51], caused by C and N (interstitial) at low temperature (< 400°C), and by Cr (substitutional) at high temperature (> 400 °C) [52]. L-PBF 316L SS has been reported to contain a trace amount of N [17] and chemical analysis of this study confirmed lower C content in contrast to standard AISI 316L SS (Supplementary Table 2) [53], making N likely to be the sole driving factor for DSA at low temperatures. Moreover, STEM-EDS over grain boundary in the non-substructured region revealed no (macro-) segregation (Supplementary Fig. 6), but the Diffusion Distance (x) of Cr was calculated to be on the order of the dislocation cell spacing, i.e. ~1 µm (Supplementary Calculation 3, [54]).

The comparatively lesser extent of dislocation density in the inclined strut, where prevalence of non-substructured (planar solidification) regions was observed, can be explained by the presence of segregated elements with equal or smaller atomic size than Fe (Cr, N from residual air in the build chamber [55] or during powder atomization [17], and possibly Si [52]) that would have lesser resistance to dislocation movement (also the degree of segregation by DSA would be to a much lower extent due to the small time frame incidence in L-PBF, i.e. likely to be few atoms, thus creating substructures undetectable in SEM and equally difficult to track in STEM) compared to larger size atoms segregated in the substructured (cellular solidification) regions. The presence of two different sets of micro-segregated elements is evident on the etched surface because of the difference in chemical erosion rate between the cell walls and their interiors, where cell walls become raised ridges and depressed grooves in substructured and non-substructured regions, respectively [42].

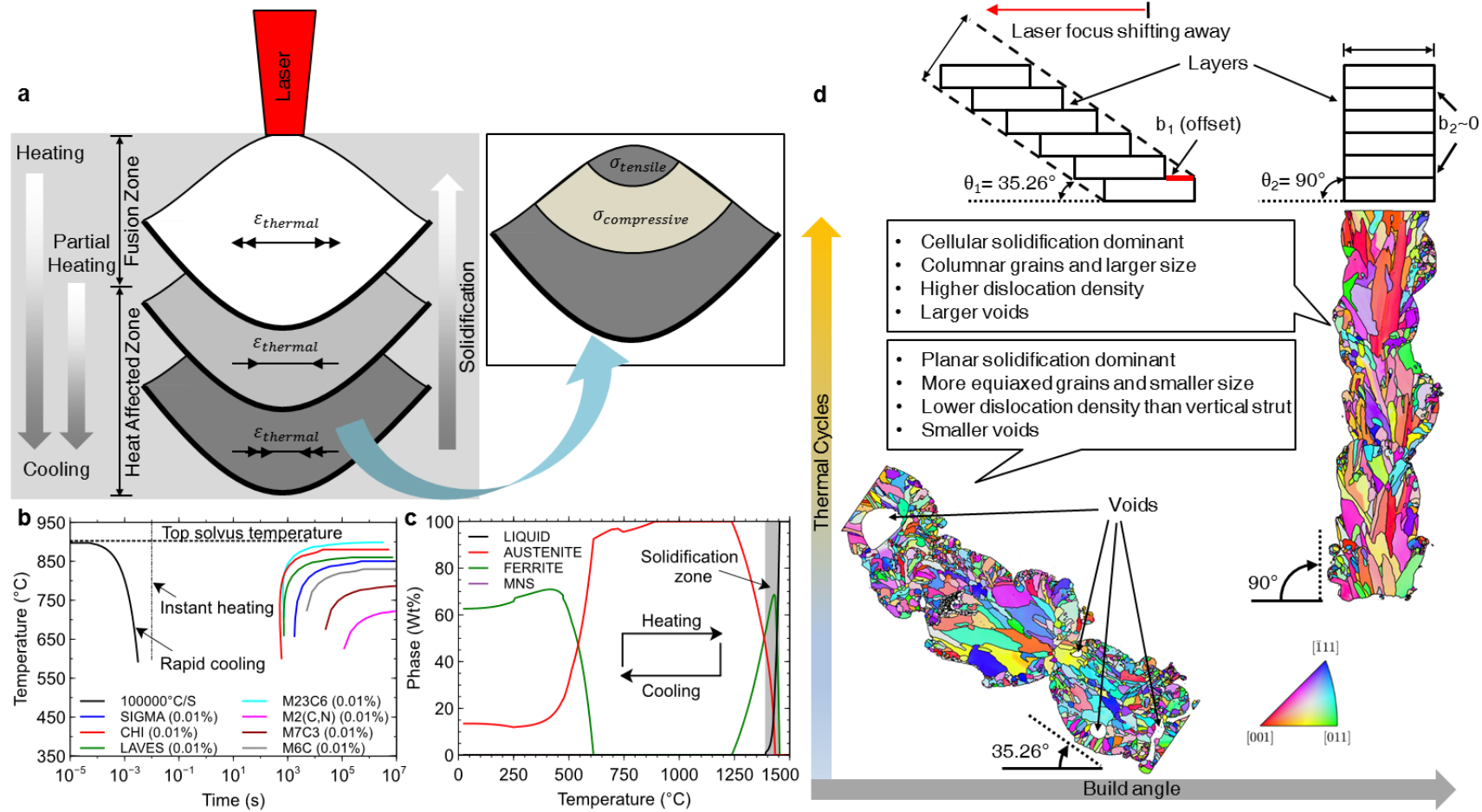

Figure 4: (a) Successive re-visitation of laser creating Fusion Zone (FZ) and Heat Affected Zone (HAZ) with respect to solidification direction by full and partial heating at the latest and prior (solidified) exposure points, respectively; (b) the rate difference in (c) repetitive heating and cooling thermal cycles. M and C in (b) stand for Metal Carbides (e.g. FeC, NiC, CrC etc.) with various ratios of M to C, (d) Effect of thermal cycles with respect to build angle.

**5.1.3 Porosity**

Individual build layers were identified with horizontal dotted lines by visual identification of the apparent position of melt pool boundaries seen on the strut outer surface (Supplementary Fig. 7) as a qualitative measure since the solidified melt pool does not remain an exact (horizontal) plane. An affinity of the voids to form away from the interface between powder layers was observed, leading to a trend where smaller voids reside closer to the layer boundaries with larger voids residing mostly inside the layer. The difference in characteristic properties of generated voids (position, size, shape etc) can be explained by how poor fusion affects PBF process. The laser parameters (energy) used to manufacture the microlattice struts in current study were optimized for maximum density [56] and the powder particle size distribution was judged to be well graded [57]. However, uniform refinement of neighbouring melt pool zones was not possible [58] since there was no "regular" (horizontal) scanning pattern as the single spot exposure strategy was used. Spatter ejection of powder and matter [59,60] accompanied by the keyhole effect [61] contributed to the major share of pore formation due to the said single exposure at the strut axis, evident from the tendency of larger voids being relatively centred on melt pool, i.e. mid cross-section (Supplementary Fig. 7). Melt defects generate during manufacturing, where a strong Marangoni-driven flow causes droplet spatter since melted 316L SS has a low viscosity, while the same flow also promotes gas release into the atmosphere, resulting in powder spatter [62]. And the defects further gets complemented by vapor-filled depressions (keyhole mode) triggered at a critical power density [61].

**5.2 Plastic Deformation Mechanism**

The plastic deformation in austenitic SS can be mediated either by dislocation glide (slip), mechanical twinning and/or martensitic transformation, and the co-existence of these mechanisms is often possible [63]. The Stacking Fault Energy (SFE), which depends on the chemical composition and temperature, dictates the active mechanism(s) of plastic deformation [64]. The SFE was calculated to be 32.94 mJ/m$^2$ according to the empirical formula given by Pickering [65], based on the chemical composition of the powder particles (Supplementary Table 2). Martensitic transformation is ruled out as it has been suggested that an SFE lower than 16-20 mJ/m$^2$ is required for it to dominate the plastic deformation mechanism [66], therefore the deformation mechanism of SLM produced 316L SS of this study would be governed by dislocation slip and further complemented by deformation twinning.

**5.2.1 Twinning**

The as-built struts exhibited a small number of grown-in twins that are "annealed" in nature (larger and straight across, not lenticular) (Supplementary Fig. 8), as a result of elevated temperature PBF thermo-mechanical processing (i.e. pseudo hot-working) experienced by 316L SS [67]. The planar solidified regions had smaller gradient in misorientations (i.e. total misorientation relative to the grain average orientation, termed as Grain Reference Orientation Deviation or GROD) within each grain (and low SFE due to addition of N [55]), while the cellular solidified regions had relatively larger gradient in misorientations among the cells (by segregation of larger atoms such as Mo) that makes it comparatively harder for the large coherent shearing event of crystal lattice (i.e. twin formation) to occur (Supplementary Fig. 9). Thus, non-substructured regions displayed more twinned boundaries than substructured regions, as seen in this study and by others [42]. As a result, the inclined strut exhibited more twinned boundaries than the vertical strut, aided by the shift from cellular to planar dominant solidification as the strut build angle deviates from the vertical orientation.

Similar trend of build angle dependent degree of twin formation was observed in the IPF map of deformed struts, both inclined (Fig. 5c and d) and vertical (Fig. 5e and f), after compression of bulk samples (at room temperature) reveal extensive formation of deformation twins in addition to annealing twins. The deformed inclined strut exhibits far more twinned grains (CSL(3) in *fcc* is 60° about <111> axis) than the deformed vertical strut, though too few coherent boundary segments were observed in all the IPF maps. However, formation of nano-twins (2-6 nm thickness) from the pre-existing cell walls, not necessarily from the grain boundaries, was observed [68] in the substructured region of SLM manufactured 316L SS. Moreover, it was reported that the previously discussed dislocation network keeps the plastic flow steady by allowing transmission of dislocations through the dislocation cell walls with stress increment (e.g. soft impediment), while the segregated atoms/particles on the walls stabilize the dislocation network by pinning mechanism (i.e. keeping the wall size intact) [68]. The amount of nano- and micro-scale twinning (causes the dynamic Hall-Petch relationship to activate by reducing mean-free paths) was governed by micro-segregation that further causes local SFE variation in the substructured and non-substructured regions, respectively.

### 5.2.2 Dynamic Recrystallization

Clusters of smaller grains were observed in post-deformation samples (compared to pre-deformation grain size) along with still nucleating grains at the HAGBs induced by localized rotation of the lattice within the crystal but near grain boundaries that would result in a necklace like feature [69], with the extent being particularly prominent in the inclined strut (Fig. 5c and d) than the vertical strut (Fig. 5e and f). This unusual occurrence of continuous dynamic recrystallization (CDRX) although at room temperature triggered by high stress (Fig. 5a and b) as seen from numerical simulation of unit cells, akin to dynamic recrystallization under Severe Plastic Deformation (SPD) [69] in a low SFE alloy such as 316L SS that was manufactured by SLM process, was previously proposed [70] to be aided by pre-existing cellular substructures (sub-grains) that may act as nuclei for CDRX. Nevertheless, the finer grain size and pre-existing dislocations (via thermal stresses) in the planar solidification regions helped to form thicker dislocation cell walls upon deformation and enhanced the conversion of LAGBs to HAGBs (i.e. CDRX). On the other hand, the soft impediment of dislocation motion by the substructure walls in the cellular solidification region inhibited the formation of thick enough dislocation walls to cause CDRX, not to mention the classical grain boundary was also far away due to larger grain size. Hence, solid-state diffusion, not solidification, induced micro-segregation triggered the unconventional CDRX. Consequently, extensive occurrence of CDRX was observed in the non-substructure (region) dominated inclined strut than substructure (region) dominated vertical strut.

### 5.2.3 Effect on Strength-Ductility

The smaller grain size in the inclined struts (Fig. 1b) would, in general, contribute to strengthening through the Hall-Petch effect. However, that was not the case for samples under current study (hardness plot in Fig. 3b). It was rather opposite because of the difference in total dislocation density (as-built or initial). The higher dislocation density in the vertical strut can be attributed to the (comparatively) slow solidification where the complex heat transfer mechanism was in effect for a longer time (subjected to more thermal cycles) than the inclined strut, following the cyclic stress reversal scheme depicted in Fig. 4a and c that caused the sample to be more pre-strained (i.e. work hardened).

The deformed inclined strut in Fig. 5c and d has smaller grains and less prominent KAM values (Supplementary Fig. 10 and 11) because it went through glide, twinning and CDRX (not much LAGBs left to convert to HAGBs [70]); which explains relatively smooth IPF orientation gradient as dislocations were absorbed [71] by newly formed grain boundaries of recrystallized grains (causing rather low dislocation density in grain interior, hence increasing ductility). The deformed vertical strut with larger grain in Fig. 5e and f did not experience enough stress to complete glide, had less amount of twinning and CDRX (LAGBs remaining to be converted to HAGBs [70]) because of the comparatively lesser stress since twinning mechanism is inversely proportional to grain size (Fig. 5b). Moreover, IPF orientation gradient was not smooth due to high initial dislocation density that could not be completely utilized (absorbed via CDRX). Additionally, dislocation glide was easier for larger grain sample as impediment was far away, this also caused twinning and/or CDRX to trigger late as the sample had more room to deform plastically by glide only. Hence rendering the vertical strut to be more ductile than the inclined strut, but pre-existing voids would be the final deciding factor for strut strength-ductility.

For the bulk structure, existing pores act as generator sites for structural discontinuities under loading (as deduced from strut tensile tests [56]) and would inhibit individual struts from attaining full strength by premature failure due to development of stress concentrations. Furthermore, BCC block deformation was more stable (i.e. uniform) than BCC-Z (Fig. 3) because of difference in dominant strut collapse mode (bending of inclined strut and buckling of vertical strut, respectively), where formation of expanding shear bands (akin to slip of dislocations in atomic crystal [7]) in BCC-Z blocks were expedited by larger voids of vertical struts. Nevertheless, the BCC-Z blocks were able to accommodate greater amount of strain by localization within a narrow band of meta-crystal unit cell(s) and resisted more load (by not having all the vertical struts buckled at once) than BCC blocks that did not exhibit this bounded (restricted) deformation mechanism. Optimization of unit cell topology and elimination (or reduction) of voids would enable the lattice structures, especially BCC-Z, to exhibit further augmented stress-strain response, and the required adjustments have to be incorporated before and during the build process [72], respectively. In addition, the process-structure-property-performance of the current study are for the novel single spot laser exposure approach and different outcome would be observed for other materials or for the conventional contour-hatch laser approach using the same material since the melt pool characteristics (e.g. size, shape, depth etc.) would be different [73].

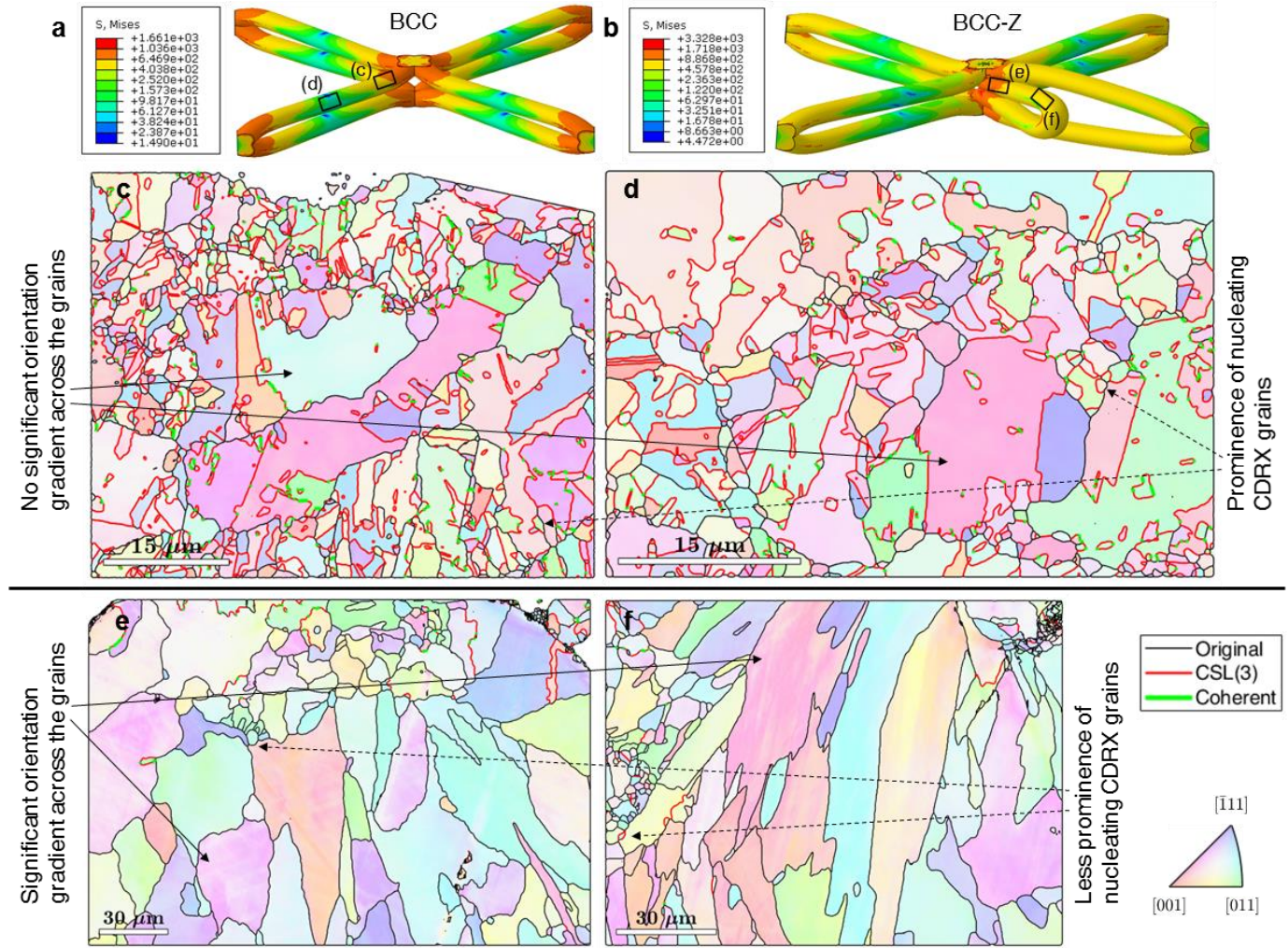

Figure 5: (a-b) FE simulation of unit cells compressed up to 80% strain, IPF maps (step size: 110 nm) of deformed struts (mid cross-section) from bulk compression samples - (c-d) inclined and (e-f) vertical strut, legend indicates the boundary colours (boundaries ≥ 2°).

# 6 Conclusions

The realization of ultralight-weight lattice structures consisting of thin struts at the micro-scale in laser PBF metal additive manufacturing was possible with the utilization of the single spot/point exposure technique. The significant physical processes involved and their influences on the 316L SS microstructure were also identified and analysed. The unexpectedly good (bulk) performance is attributed to the complex thermo-mechanical processing during the build process (successive heating-cooling cycles in the melt pool) that acted as pseudo hot-working and made the as-built struts work-hardened (through the generation of dislocation structures), thus increasing the yield strength (~ 2.7 to 3.1 times that of annealed condition, depending on the build angle). On the other hand, instead of exhibiting decreased ductility, as is the norm with work-hardened metals/alloys, the 3D printed material showed increased ductility because of substructures generated from solute-segregation. The combined effect of rapid heating-cooling cycles at high temperature gradients due to the variation in build angle that governs the size of heat affected zone (HAZ) relative to fusion zone (FZ), thereby controlling the complex solidification path and the resultant residual stress beyond yield point as a result of thermal stresses, dictates micro-segregation by either solidification or strain aging mechanism (i.e. solid-state diffusion). This enables L-PBF manufactured 316L stainless steel to achieve high toughness, eliminating the need for heat-treatment since as-built parts could possibly perform better, and renders its ultralight-weight microlattice structural form a suitable candidate for impact/blast resistance applications (e.g. defence/automotive/aerospace industry). Moreover, the use of two different unit cells and their comparative study demonstrated the usefulness of strain localization mechanism in bulk deformation of architected materials and the possibility of enhancing, or better yet, controlling the meta-atomic bonds (i.e. strut strength-ductility) by harnessing the non-equilibrium solidification in additive manufacturing. Overall, this work gives a detailed analysis of microlattice strut microstructure and bulk mechanical properties for a specific set of laser parameters, and paves the way for manufacturing high performing lattice structures (e.g. single meta-grains) of different metals/alloys with desired strength-ductility balance requirement where the build parameters can be adjusted to vary and tailor to target strut properties, that would further enable the fabrication of damage-tolerant architected hierarchical cellular materials by utilizing optimized macro-crystal topologies in conjunction with other hardening principles of metallurgy (e.g. grain boundaries, precipitates, and phases).

## Acknowledgement

The authors thankfully acknowledge the access to Electron microscopy/experimental facilities at ANSTO (LEX2823 - Student agreement), and tomography facility in the National Laboratory for X-ray Micro Computed Tomography (CTLab) at Australian National University and Benjamin Young at Thermo Fisher Scientific. The authors thank LPW Technology® (UK) for conducting the chemical and particle size analysis on the 316L stainless steel powder. Moreover, the authors are grateful for the support of Prof. Jodie Bradby and Ms. Xingshuo Huang in the Department of Electronic Materials Engineering at the Australian National University in facilitating the nanoindentation measurements. Finally, the authors extend sincere appreciation to Dr. Sherry Mayo for tomography advice, Dr. Dierk Raabe for having discussions related to this work, Dr. Sharvan Kumar and Dr. Allan Bower for providing original copy of Fig. 3e; Dr. Henrik Kaker, Dr. Juan Pablo Escobedo-Diaz and Dr. Safat Al-Deen for technical help.

## Author contributions

MGR conceived the project. M Smith and RAWM manufactured the microlattice samples and conducted bulk compression tests. MS and MGR carried out the Nano-CT data acquisition, reconstruction and processing/analysis of the reconstructed image stacks. MGR carried out Finite Element Analysis, Nanoindentation experiments and the produced data processing/analysis. DB designed and performed the electron microscopy (BSE/EBSD/EDS/TEM) along with AX and MGR, MGR performed the resulting data processing/analysis, formulation, code writing and results interpretation. MGR performed the paper writing including all illustrations, with feedback from all other authors led by DB.

## Competing interests

The authors declare no competing financial interests.

## Data availability

Representative samples of the research data are given in the figures (and supplementary data). Other datasets generated and/or analysed during this study are not publicly available because those are part of ongoing research.